\documentclass[twocolumn,showpacs,prb,amsmath,amssymb,superscriptaddress]{revtex4-1}

\usepackage{color}
\usepackage{epstopdf}
\usepackage{graphicx}
\usepackage{dcolumn}
\usepackage{bm}
\DeclareFontFamily{U}{rsfs}{\skewchar\font127}
\DeclareFontShape{U}{rsfs}{m}{n}{<-6>rsfs5<6-8.5>rsfs7<8.5->rsfs10}{}
\DeclareSymbolFont{rsfs}{U}{rsfs}{m}{n}
\DeclareSymbolFontAlphabet{\mathrsfs}{rsfs}
\DeclareRobustCommand*\rsfs{\@fontswitch\relax\mathrsfs}

\begin{document}

\preprint{APS/123-QED}

\title{Localize and mixed valence states of Ce $4f$ in superconducting and ferromagnetic
 CeO$_{\bm{1-x}}$F$_{\bm{x}}$BiS$_{\bm{2}}$ 
revealed by  x-ray absorption and photoemission spectroscopy}

\author{T. Sugimoto}
\author{D. Ootsuki}
\affiliation{Department of Physics \& Complexity Science and Engineering, University of Tokyo, 5-1-5 Kashiwanoha 277-8561, Japan}
\affiliation{Institute for Solid State Physics, University of Tokyo, 5-1-5 Kashiwanoha 277-8561, Japan}

\author{E. Paris}
\affiliation{Dipartimento di Fisica, Universit\'a di Roma ``La Sapienza'' - Piazzale Aldo Moro 2, 00185 Roma, Italy}
\affiliation{Center for Life NanoScience@Sapienza, Istituto Italiano di Tecnologia, V .le Regina Elena 291, 00185 Rome, Italy}

\author{A. Iadecola}
\affiliation{ESRF - The European Synchrotron, 71 avenue des Martyrs, 38043 Grenoble Cedex 9, France}

\author{M. Salome}
\affiliation{ESRF - The European Synchrotron, 71 avenue des Martyrs, 38043 Grenoble Cedex 9, France}

\author{E. F. Schwier}
\author{H. Iwasawa}
\author{K.~Shimada}
\affiliation{Hiroshima Synchrotron Radiation Center, Hiroshima University, Hiroshima 739-0046, Japan}

\author{T. Asano}
\author{R. Higashinaka}
\author{T. D. Matsuda}
\author{Y. Aoki}
\affiliation{Department of Physics, Tokyo Metropolitan University, Tokyo 192-0397, Japan}

\author{N. L. Saini}
\affiliation{Dipartimento di Fisica, Universit\'a di Roma ``La Sapienza'' - Piazzale Aldo Moro 2, 00185 Roma, Italy}

\author{T. Mizokawa}
\affiliation{Department of Applied Physics, Waseda University, Tokyo 169-8555, Japan}

\date{\today}

\begin{abstract}
We have performed Ce $L_3$-edge x-ray absorption spectroscopy (XAS) and Ce $4d$-$4f$ resonant photoemission spectroscopy (PES) on single crystals of CeO$_{1-x}$F$_x$BiS$_2$ for $x=0.0$ and 0.5 in order to investigate the Ce $4f$ electronic states. In the Ce $L_3$-edge XAS, mixed valence of Ce was found in the $x=0.0$ sample and the F-doping suppresses it, which is consistent with the results on polycrystalline samples. As for the resonant PES, we found that the Ce $4f$ electrons in both $x=0.0$ and $0.5$ systems respectively form a flat band at 1.0 eV and 1.4 eV below the Fermi level and there is no contribution to the Fermi surfaces. 
Interestingly, Ce valence in CeOBiS$_2$ deviates from Ce$^{3+}$ even though Ce $4f$ electrons are localized, indicating the Ce valence is not in a typical valence fluctuation regime.
We assume that localized Ce $4f$ in CeOBiS$_2$ is mixed with the unoccupied Bi $6p_z$, which is consistent with the previous local structural study. Based on the analysis of the Ce $L_3$-edge XAS spectra using Anderson's impurity model calculation, we found that the transfer integral becomes smaller increasing the number of Ce $4f$ electrons upon the F substitution for O.
\end{abstract}

\maketitle

\newpage

Since the discovery of superconductivity in the BiS$_2$ system by Mizuguchi {\it et al.} \cite{Mizuguchi2012}, the electronic and lattice structures of various BiS$_2$-based superconductors have been attracting great interest including REO$_{1-x}$F$_x$BiS$_2$ (RE: rare earth elements)\cite{2,3,4,5,6,7,8,9,10,11,12,13,14,15,16}. In a typical REO$_{1-x}$F$_x$BiS$_2$ system, REO$_{1-x}$F$_x$ block layer and BiS$_2$ layer are alternatively stacked each other, and F substitution for O is considered as an electron dope to the electronically active BiS plane \cite{Morice2013} whereas it has continuously been found that actual carrier concentration is always smaller than its nominal value \cite{Ye2014,Zeng2014,Sugimoto2015}. Among the various REO$_{1-x}$F$_x$BiS$_2$ compounds, CeO$_{1-x}$F$_x$BiS$_2$ is very unique in that the superconductivity in the BiS$_2$ layer and the ferromagnetism in the CeO$_{1-x}$F$_x$ layer coexist for $x > 0.4$ \cite{Demura2015}.
Meanwhile, an x-ray absorption spectroscopy (XAS) study on  CeO$_{1-x}$F$_x$BiS$_2$ has revealed that the Ce valence is intermediate (valence fluctuation) and the Ce $4f$ states may contribute to the Fermi surfaces for $x<0.4$, and the F-doping makes the system crossover from valence fluctuation regime to Kondo-like regime \cite{Sugimoto2014}. However, from the DC susceptibility measurement on  CeOBiS$_2$, Ce $4f$ electrons are found to be in a well-localized state \cite{Higashinaka2015}, which is inconsistent with the XAS results. In order to identify the role of the Ce $4f$ electrons, we have investigated the electronic states of Ce $4f$ in CeO$_{1-x}$F$_x$BiS$_2$ using Ce $L_3$-edge XAS and Ce $4d$-$4f$ resonant ARPES.


High-quality single crystals of CeOBiS$_2$ and CeO$_{0.5}$F$_{0.5}$BiS$_2$ have been prepared by CsCl flux method \cite{Higashinaka2015}. As for the stoichiometry of $x=0.5$ sample, the observed lattice constant $c=13.443$\AA   is reasonable compared with previous studies on Ce(O,F)BiS$_2$ system \cite{Demura2015,Nagao2014}.
The Ce $L_3$ XAS measurements were performed at ID21, ESRF. At ID21, a double-crystal fixed-exit Si(111) monochromator (Kohzu, Japan) was used for energy scans. The beam was focused down to a micro-probe by a Kirkpatrick-Baez mirrors system.
All the spectra were taken in fluorescence yield mode, the fluorescence detector was a 10 mm$^2$ Silicon drift diode (R\"{o}ntec, Germany). Acquisition was performed at low temperature using a liquid nitrogen cryostat.
The resonant ARPES measurements with linearly polarized photons were performed at the undulator beamline BL-1 of HSRC, Hiroshima University. The endstation is equipped with Scienta Omicron R4000 analyzer. The photon energy was set to be 30 - 130 eV. We cleaved the single crystalline sample \textit{in situ} under ultrahigh vacuum ($<1\times10^{-10}$ Torr) to obtain clean (001) surface. The total energy resolutions were measured to be 21 meV and 103 meV at $h\nu=$30 eV and $h\nu=$120 eV, respectively. The angular resolution was 0.7 deg., corresponding to 0.032 \AA$^{-1}$ for 30 eV and 0.067 \AA$^{-1}$ for 130 eV in momentum space. All the ARPES measurements were performed at 50 K with $p$-polarization geometry.



\begin{figure}
\includegraphics[width=7cm]{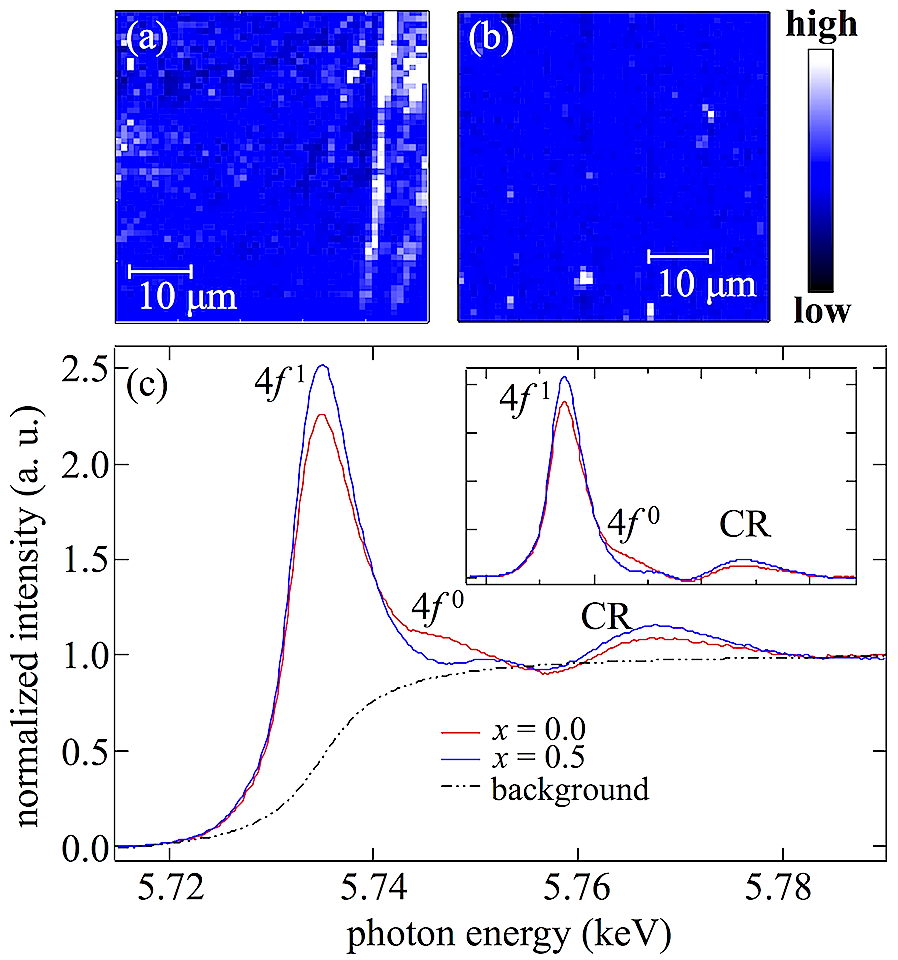}
\caption{(Color online) (a)  Real space image of $\mu$XANES at Ce $L_3$-edge (5.732 keV) on CeOBiS$_2$ (b) and on CeO$_{0.5}$F$_{0.5}$BiS$_2$. (c) Normalized XAS spectra of CeOBiS$_2$ and CeO$_{0.5}$F$_{0.5}$BiS$_2$. The background (arctangent function) is denoted by the dot-dashed line. The spectra after background subtraction are also shown in the inset.
}
\label{fig1}
\end{figure}

Figures 1(a) and 1(b) show the real space 50 $\mu$m $\times$ 50 $\mu$m images of XAS intensities at Ce $L_3$-edge (5.732 keV) on CeOBiS$_2$ and CeO$_{0.5}$F$_{0.5}$BiS$_2$, respectively. From these images, one can see that the Ce $4f$ electronic states are spatially homogeneous in this scale. We also confirmed that the spectral shape at the dark spot is the same as that of the bright spot, indicating that the intensity modulation is not derived from electronic inhomogeneity.

Figure 1(c) shows the normalized Ce $L_3$-edge XAS spectra of CeOBiS$_2$ and CeO$_{0.5}$F$_{0.5}$BiS$_2$. 
The spectra were normalized with respect to the atomic absorption estimated by a linear fit to the high energy part of the spectra as used in the earlier study \cite{Sugimoto2014}.
The background (arctangent function) is shown by the dot-dashed line. The background-subtracted spectra are also shown in the inset with the same energy and intensity scales.
Three main structures around 5.732, 5.745, and 5.768 keV can be identified in the Ce $L_3$-edge XAS spectra. The first (second) feature is the transition from Ce $2p$ to Ce $5d$ mixed with $4f^1$ ($4f^0$) final state \cite{Thole1985,Matsuyama1997,Kotani2012,Kroll2009}. The coexistence of the first and second peaks suggests that Ce$^{3+}$ and Ce$^{4+}$ coexist in the system. Here, the spatially homogeneous XAS map indicates that the observed $f^0/f^1$ mixed valence is not due to CeO$_2$ impurities but is an intrinsic physics. The third feature is so-called continuum resonance (CR) derived from the Ce-Bi scattering, and reflects the Ce-Bi bondlength \cite{Sugimoto2014}. In addition to the three main features, there is a weak structure around 5.75 keV. This is the typical structure of layered rare-earth system, and its intensity is sensitive to the O/F atomic order/disorder in Ce(O,F) layer \cite{Sugimoto2014}. All the XAS results are quantitative consistent with the previous Ce $L_3$-edge XAS results of polycrystalline CeO$_{1-x}$F$_{x}$BiS$_2$ \cite{Sugimoto2014}. 
However, the inconsistency with the DC susceptibility measurement still remains; in CeOBiS$_2$, the XAS result suggests that the Ce $4f$ electrons are valence fluctuating whereas the DC susceptibility measurement found Ce $4f$ well-localized. We have performed Ce $4d$-$4f$ resonant ARPES on the CeOBiS$_2$ and CeO$_{0.5}$F$_{0.5}$BiS$_2$ single crystals in order to investigate the Ce $4f$ electronic states and specify the orbitals mixing with the Ce $4f$ in CeOBiS$_2$.

\begin{figure}
\includegraphics[width=8cm]{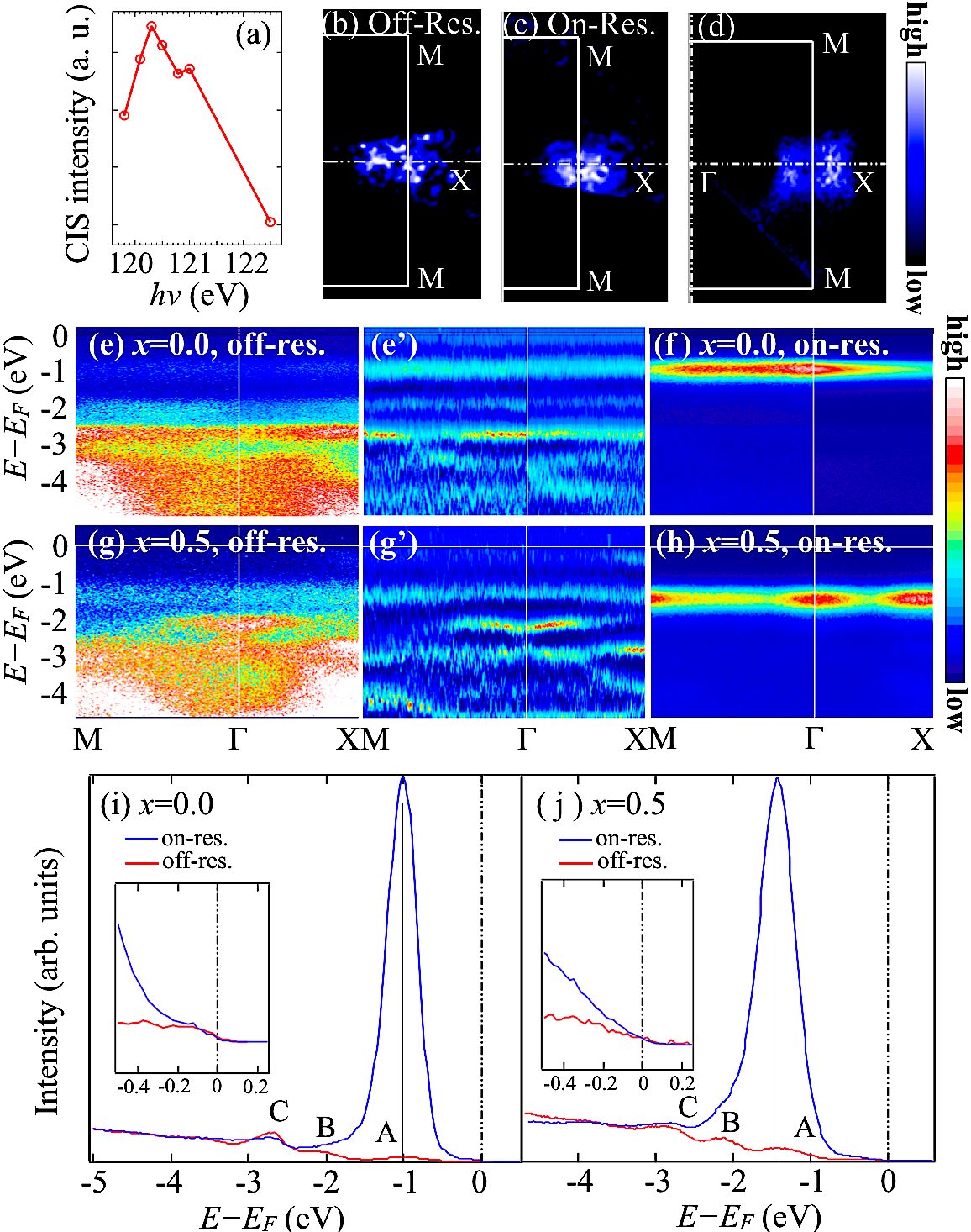}
\caption{(Color online) (a) CIS plot of CeO$_{0.5}$F$_{0.5}$BiS$_2$.  Fermi surfaces of CeO$_{0.5}$F$_{0.5}$BiS$_2$ taken (b) at on-resonant 120.3 eV, (c) at off-resonant 115.4 eV, and (d) at 30 eV. The ARPES intensities are integrated within $\pm$ 50 meV with respect to $E_F$. (e)  Cut along M-$\Gamma$-X  with off-resonant $h\nu$ and (e') its second derivative,  (f) cut with on-resonant photon energy,  (i) EDCs integrated along M-$\Gamma$-X for both off- and on-resonant $h\nu$, and the inset shows the expanded view near $E_F$ on CeOBiS$_2$. (g), (g'), (h), and (j) are the same as (e), (e'), (d), and (i) but on CeO$_{0.5}$F$_{0.5}$BiS$_2$.
}
\label{fig2}
\end{figure}

Figure \ref{fig2}(a) shows constant initial state (CIS) plot with respect to photon energy at the Ce 4$f$ peak of CeO$_{0.5}$F$_{0.5}$BiS$_2$ normalized by photon flux, which determines the Ce 4$d$-4$f$ on-resonant photon energy of the system as 120.3 eV. Figures \ref{fig2}(b) and \ref{fig2}(c) show the Fermi surface maps of CeO$_{0.5}$F$_{0.5}$BiS$_2$ at off-resonant 115.4 eV and on-resonant 120.3 eV. They are obtained by rotating the azimuthal angle with respect to the crystallographic $c$ axis of the sample. Since there is no considerable changes in Fermi surfaces with the on- and off-resonant photon energies, the Ce $4f$ electrons are not employed as a part of the Fermi surfaces.  The Fermi surfaces measured at 30 eV are shown in Fig. \ref{fig2}(d), whose Luttinger volume 0.22 whereas the nominal $x$ is 0.5 (see detail in Ref. \cite{Sugimoto2015}).

Figures \ref{fig2}(e) and \ref{fig2}(f) (Figs. \ref{fig2}(g) and \ref{fig2}(h)) show the ARPES data taken at 115.4 eV (off-resonance) and 120.3 eV (on-resonance), respectively, for CeOBiS$_2$ (for CeO$_{0.5}$F$_{0.5}$BiS$_2$) along M-$\Gamma$-X. Figures \ref{fig2}(i) and \ref{fig2}(j) respectively show the energy distribution curves (EDCs) integrated along M-$\Gamma$-X taken for CeOBiS$_2$ and CeO$_{0.5}$F$_{0.5}$BiS$_2$ at on- and off-resonant photon energies, and the inset shows the expansion near $E_F$. One can see from the EDCs [Figs. \ref{fig2}(i) and \ref{fig2}(j)] that the structures A, B, and C of CeO$_{0.5}$F$_{0.5}$BiS$_2$ are roughly shifted to the higher binding energy by 0.4 eV due to the electron doping compared with those of CeOBiS$_2$, but one can also see there is a slight band-dependence of the shifts, which could be due to local atomic structural changes \cite{Paris2014}.
The on-resonance spectra are dominated by the partial density of states (DOS) of Ce 4$f$ due to the resonant enhancement. If the system is in a conventional valence fluctuating regime, the on-resonant spectrum considerably enhances two features; at approximately 0 eV ($f^0$ component) due to a strong $cf$ hybridization and at $-1.0$ eV ($f^1$ component) in CeOBiS$_2$ [at 0 eV ($f^0$) and at $-1.4$ eV ($f^1$) in CeO$_{0.5}$F$_{0.5}$BiS$_2$] with respect to $E_F$. On the other hand, if the system is in a Kondo-like regime, the on-resonant enhancement can be seen in one feature; at $-1.0$ eV in CeOBiS$_2$ (at $-1.4$ eV in CeO$_{0.5}$F$_{0.5}$BiS$_2$) with respect to $E_F$, which is $f^1$ component  \cite{Gunnarsson1983}. 

In CeO$_{0.5}$F$_{0.5}$BiS$_2$, the Bi $6p$ Fermi pocket around X point is observed in both the on- and off-resonance data. The valence band below $-1$ eV can be assigned to Bi $6s$, S $3p$, and Ce $4f$ bands. When the photon energy is tuned to the Ce $4d$-$4f$ absorption energy (120.7 eV), the flat band located at $-1.4$ eV is considerably enhanced. 
The resonant ARPES results show that the Ce $4f$ band is located between the Bi $6p$ conduction band and the Bi $6s$/S $3p$ valence band, and that there is no appreciable Ce $4f$ spectral weight at $E_F$. Namely, the results indicate that the Ce $4f$ electrons are basically localized (Kondo-like regime), do not contribute to the Fermi pockets around X, and its valance should be $3+$. This result on CeO$_{0.5}$F$_{0.5}$BiS$_2$ is consistent with the bulk sensitive Ce $L_3$-edge XAS result where CeOBiS$_2$ falls in the valence fluctuation regime and gradually changes to the Kondo-like regime with the F substitution for O \cite{Sugimoto2014}. The superexchange or RKKY interaction between the localized Ce $4f$ moments is derived from the Ce $4f$-S $3p$ hybridization.

On the other hand, in CeOBiS$_2$, we found that the Ce $4f$ lineshape is very similar to that of CeO$_{0.5}$F$_{0.5}$BiS$_2$ as shown in Figs. \ref{fig2}(i) and \ref{fig2}(j), and is different from what one expected for valence fluctuating systems. In typical valence fluctuating systems such as CeRu$_2$, the Ce $4f$ band is strongly hybridized with the conduction band such as Ru $4d$ and is well reproduced by local density approximation (LDA) band calculations. The observed Ce $4f$ band of CeOBiS$_2$ has very flat dispersion and does not agree with the LDA band calculation \cite{Sugimoto2015}. 
Here we speculate that the Ce $4f$ electrons in CeOBiS$_2$ are mixed with the unoccupied Bi $6p_z$. From the crystal structure, the nearest ion for Ce is oxygen, but the oxygen band is located far from $E_F$ by 3 - 4 eV and therefore hard to hybridize with the Ce $4f$. The other candidate for the mixing partner of the Ce $4f$ is the unoccupied Bi $6p_z$ through the out-of-plane S. 
Since the Bi $6p_z$ orbitals do not contribute to the Fermi surfaces, one electron in the Ce $4f$ orbitals should be completely transferred to the Bi $6p_z$ orbitals in order to obtain the $f^0$ component in the Ce $L_3$ XAS. However, if the electron transfer from the Bi $6s$ orbitals to the Ce $4f$ orbitals is included, the mixture of the $f^1$ and $f^0$ states becomes possible and the Ce $L_3$ XAS result would be explained. 
Note that the Ce $4f$-Bi $6p_z$ hybridization does not necessarily leads to a metallic state. It provides the mixed valence state given by the linear combination of $|4f^1\rangle$ and $|4f^0\; L^1\rangle$ where $L$ is Bi $6p_z$. This state can be localized if the overlap between the neighboring Bi $6p_z$ states is sufficiently small compared with the intersite Coulomb repulsive energy. Even if the overlap is not small enough for the localization due to the pure Coulomb interaction, the combination of the Coulomb interaction and the atomic disorder \cite{Paris2014} can provide an insulating ground state, possibly a sort of charge glass of Bi $6p_z$ electrons. Moreover, this hybridization may impede the ferromagnetic order in the Ce(O,F) layer.


In order to examine this possibility, we have performed Anderson's impurity model (AIM) calculation on Ce $L_3$-edge XAS results
 under the condition that the Ce $4f$ is mixing with the unoccupied Bi $6p_z$ and the occupied Bi $6s$.
The model Hamiltonian is given by the standard Anderson Hamiltonian including the Ce $4f$/$5d$ electrons, the conduction-band electrons, and the Ce $2p$ core-level electrons.
\begin{align}
\mathrsfs{H}_{\text{A}} =&\sum_k \varepsilon_k c^{\dagger}_k c_k + \varepsilon_f \sum_{m} f^{\dagger}_{m}f_{m} + U_f \sum_{m'>m} f^{\dagger}_{m'}f_{m'}f^{\dagger}_{m}f_{m}  \notag \\
&+ n_{\text c} \varepsilon_{\text c} +(1-n_{\text c})Q_f\sum_{m} f^{\dagger}_{m}f_{m}  \notag \\
&+ \sum_{mk} ( V_{mk} c_k^{\dagger}f_{m} + V^*_{mk}f_{m}^{\dagger}c_k) \notag \\
&  +\varepsilon_d \sum_{m} d^{\dagger}_{m}d_{m} + U_{fd} \sum_{m'>m} f^{\dagger}_{m'}f_{m'}d^{\dagger}_{m}d_{m} \notag \\
& + (1-n_{\text c})Q_d\sum_{m} d^{\dagger}_{m}d_{m},
\end{align}
Here, $f^{\dagger}_m$ ($d^{\dagger}_m$) are creation operators for the Ce $4f$ ($5d$) electrons with orbital notation $m$, and  $c^{\dagger}_k$ are creation operators for Bloch electrons in the Bi $6p_z$ and Bi $6s$ bands with wave vector $\bm{k}$.  The parameters $U_f$ and $Q_f$ ($Q_d$) are the on-site repulsive Coulomb interaction between the Ce $4f$ electrons and the attractive Coulomb interaction between the Ce $4f$ ($5d$) electron and the Ce $2p$ core hole, and $U_{fd}$ is the on-site repulsive Coulomb interaction between the Ce $4f$ and Ce $5d$ electrons. Note that the notations $m$ and $k$ include spin. The parameter $V$ describes the transfer integral between the Ce $4f$ and conduction-band electrons, $\varepsilon_f$ ($\varepsilon_d$) is the energy level of Ce $4f$ ($5d$) electrons relative to the Fermi level, and $\varepsilon_{\text c}$/$n_{\text c}$ represent the energy/number of the Ce $2p$ core-level electron.

In the Ce $L_3$-edge XAS process, the absorption occurs from the Ce $2p$ core level to the Ce $5d$ unoccupied states mixed with the Ce $4f$ states. In the framework of the Anderson's impurity model, the initial state can be simply given by
\begin{equation}
|\psi_i\rangle = \alpha|4f^0\; L^1\rangle + \beta |4f^1\rangle + \gamma |4f^2 \; \underline{L}^1\rangle
\end{equation}
where $L$ and $\underline{L}$ denote a Bi $6p_z$ electron and Bi $6s$ hole, respectively.  The final states are given by
\begin{align}
|\psi_f\rangle = \;&\alpha ' | \underline{c} \; 4f^0 \; L^1 \; 5d^1  \rangle + \beta ' |\underline{c} \;4f^1 \; 5d^1 \rangle \\ \notag
&+ \gamma ' | \underline{c} \; 4f^2 \; \underline{L}^1\;5d^1 \rangle 
\end{align}
where $\underline{c}$ denotes the Ce $2p$ core hole.
Using $\alpha$, $\beta$, $\gamma$, $\alpha'$, $\beta'$, and $\gamma'$, the spectral weight is given by $|\alpha\alpha' + \beta\beta'+\gamma\gamma'|^2$. The energy of the $| \underline{c} \; 4f^0 \; L^1 \; 5d^1 \rangle$ state is given by $\varepsilon_d-Q_d$, that of the $|\underline{c} \;4f^1 \; 5d^1  \rangle$ is given by $\varepsilon_f+\varepsilon_d-Q_f -Q_d +U_{fd}$, and that of $| \underline{c} \; 4f^2 \; \underline{L}^1 \; 5d^1 \rangle$ is given by $2\varepsilon_f+\varepsilon_d-2Q_f-Q_d+U_f+2U_{fd}$. 

The parameters $U_{fd}$ and $Q_d$ are neglected in the present analysis because they are substantially small compared with the other parameters. Since $\varepsilon_f$ can be determined by the Ce $4d$-$4f$ resonant ARPES results and $Q_f$ is fixed at $U_f/0.8$ \cite{Okada1989,Bocquet1992}, the present analysis includes two free parameters; $U_f$ and $T$, where
the effective transfer integral $T = \sqrt{\sum_{mk} \mid V_{mk} \mid^2}/\sqrt{N}$ with $N$ = 14 (see Fig. 4(a))  represents the off diagonal term.
For simplicity, the effective transfer integral is the same for the Bi 6$p_z$ and 6$s$ orbitals.

\begin{figure}
\includegraphics[width=8.5cm]{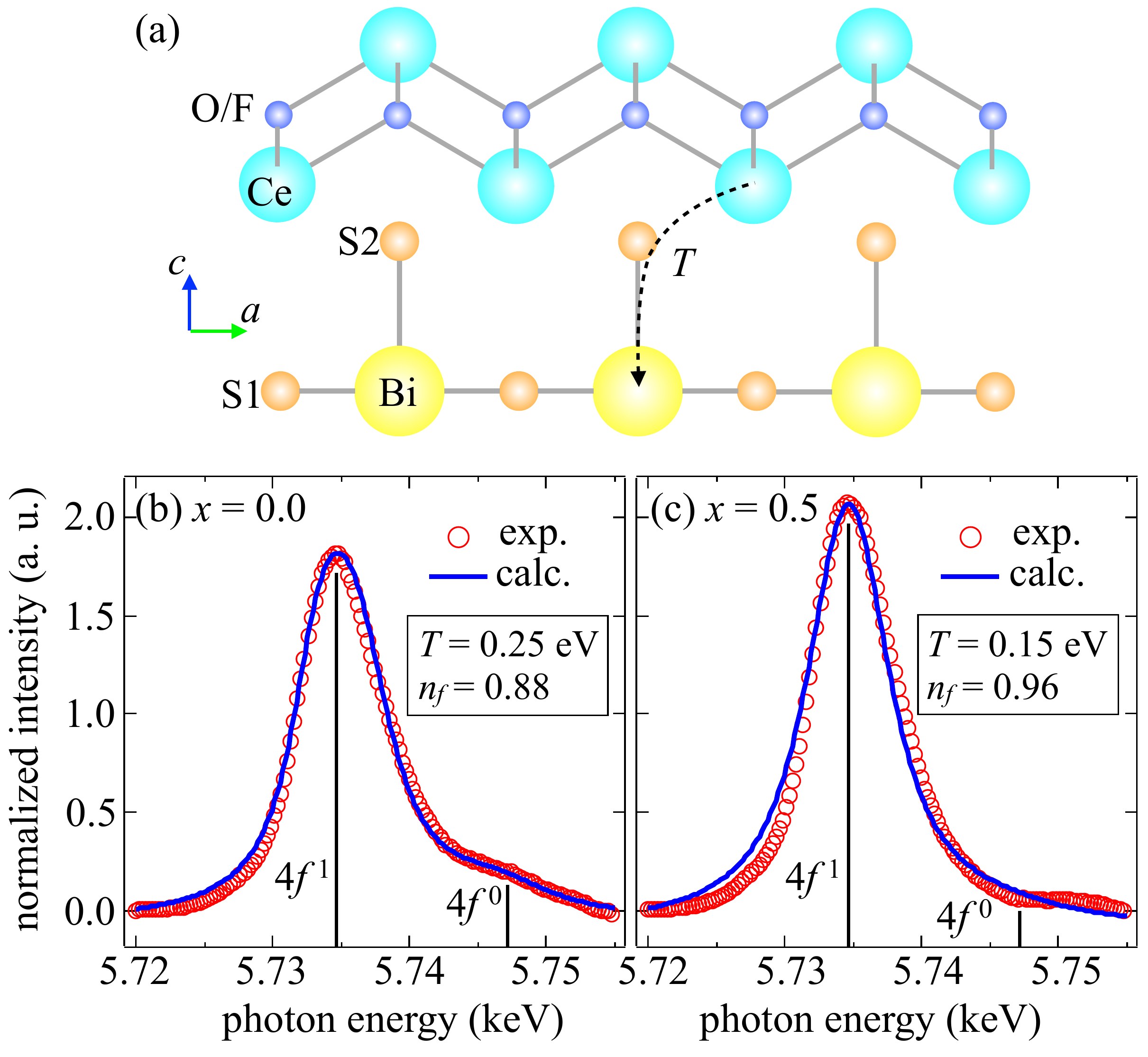}
\caption{(Color online) (a) Schematic diagram of the effective transfer integral $T$. Ce $L_3$-edge XAS experimental results on single crystal CeO$_{1-x}$F$_x$BiS$_2$  and corresponding impurity model calculations at (a) $x=0.0$ and (b) $x=0.5$. The energy positions of $4f^1$ and $4f^0$ are shown by vertical lines. The effective transfer integral ($T$) and number of Ce $4f$ electrons ($n_f$) estimated by Anderson's impurity-model calculation are also shown here.
}
\label{fig3}
\end{figure}

Considering the local structural changes of Ce, S2 (see Fig. \ref{fig3}(a)), and Bi due to the F-doping \cite{Paris2014}, the parameter $T$ should be the only adjustable parameter to reproduce the XAS spectra. Therefore, the parameter $U_f$ is fixed at 7.8 eV for all the calculations. We can reproduce the $4f^1$ and $4f^0$ structures of the XAS spectra, indicating that the scenario above could be possible. The results are shown in Figs. \ref{fig3}(b) and \ref{fig3}(c) for $x=0.0$ and 0.5, respectively, and the energy positions of the $4f^1$ and $4f^0$ structures are also shown here.
The circles and solid lines are respectively experimental and calculated results, and the evaluated $T$ and number of Ce 4$f$ electrons ($n_f$) are also shown in Fig. \ref{fig3}. The parameter $T$ changes from 0.25 eV to 0.15  eV when the F concentration $x$ goes from 0.0 to 0.5. The number $n_f$ is calculated as 0.88 and 0.96 for $x=0.0$ and 0.5, respectively. Present results are qualitatively consistent with the previous impurity model calculations on poly crystalline system \cite{Sugimoto2015a}. Moreover, this is also consistent with the thermodynamical quantities because Higashinaka \emph{et al.} reported that the amount of Ce$^{3+}$ ($4f^1$ state) in CeOBiS$_2$ is 10 \%  lower than expected \cite{Higashinaka2015}, and the impurity model calculation shows $n_f=0.88$. 

Our scenario could explain the XAS and resonant ARPES results naturally with a support of AIM calculation; the unoccupied Bi $6p_z$ states partially trap Ce $4f$ electrons and form a sort of charge glass state due to the strong atomic disorder of BiS plane.  It is expected that the electrons in the Bi $6p_z$ state are randomly distributed in real space and appear as a broad feature in the ARPES spectra. Therefore, it would rather difficult to identify the Bi $6p_z$ state from the comparison between the photoemission results and some calculations considering the atomic disorder in the BiS plane. Note that our argument is not sensitive to the value of $x$ itself as long as the system shows the superconductivity even though the uncertainty of $x=0.5$ crystal still remains here.

In summary, we have combined Ce $L_3$-edge XAS, Ce $4d$-$4f$ resonant ARPES, and AIM calculation on single crystals of CeO$_{1-x}$F$_x$BiS$_2$ for $x=0.0$ and 0.5 in order to investigate the electronic states of Ce $4f$ electrons. The XAS found that the $x=0.0$ sample is in mixed valence state that is suppressed by the F-doping, which is consistent with previous study \cite{Sugimoto2014}. The resonant ARPES found that the Ce $4f$ electrons in both $x=0.0$ and $0.5$ systems are essentially localized and there is no contribution to the Fermi surfaces. We assume that the localized Ce $4f$ in CeOBiS$_2$ is mixed with the unoccupied Bi $6p_z$ instead of Bi $6p_x/6p_y$. This scenario is consistent with the previous local structural study. The AIM calculation not only supports the scenario, but also is consistent with the thermodynamical experiments \cite{Higashinaka2015}. The hybridization of Ce $4f$ and Bi $6p_z$ may impede the ferromagnetic order in the Ce(O,F) layer.

The authors are grateful to M. Zheng and Y. Kojima for the support during the experimental run at BL-1, Hiroshima Synchrotron Radiation Center. T.S. and D.O. acknowledge the support from JSPS Research Fellowship for Young Scientists. This work is partly supported by JSPS KAKENHI (Grant No. 15H03693 and 25400356). 
E.F.S. acknowledges financial support from the JSPS postdoctoral fellowship for overseas researchers as well as the Alexander von Humboldt Foundation (Grant No. P13783). 
The synchrotron radiation experiments have been done with the approval of Hiroshima Synchrotron Radiation Center (Proposal No.14-B-24).

\end{document}